\documentclass[preprint,5p,times,twocolumn,12pt]{elsarticle-mod}

\usepackage{ifpdf}
\usepackage{graphicx,natbib,lineno,color}
\ifpdf
\usepackage{amssymb}
\usepackage{lipsum}
\usepackage{bm}
\usepackage{amssymb}
\usepackage{pifont}
\usepackage{slashed}
\usepackage{color}

\usepackage[%
breaklinks=true,%
colorlinks=true,%
linkcolor=blue,anchorcolor=blue,%
citecolor=blue,filecolor=blue,%
menucolor=blue,pagecolor=blue,%
urlcolor=blue]{hyperref}

\makeatletter
\def\elsartstyle{%
	\def\normalsize{\@setfontsize\normalsize\@xiipt{14.5}}
	\def\small{\@setfontsize\small\@xipt{13.6}}
	\let\footnotesize=\small
	\def\large{\@setfontsize\large\@xivpt{18}}
	\def\Large{\@setfontsize\Large\@xviipt{22}}
	\skip\@mpfootins = 18\p@ \@plus 2\p@
	\normalsize
}
\@ifundefined{square}{}{\let\Box\square}
\makeatother

\definecolor{purple}{rgb}{1,0,1}

\journal{ }   
\def\date{ } 

\begin{document}
	
\begin{frontmatter}
	
\title{{\bf Vorticity in analogue spacetimes}}

\author[sissa,infn]{Stefano Liberati}
\ead{liberati@sissa.it} 

\author[vuw]{Sebastian Schuster\,}
\ead{ sebastian.schuster@sms.vuw.ac.nz}

\author[sissa,infn]{Giovanni Tricella}
\ead{gtricell@sissa.it}

\author[vuw]{Matt Visser\,}
\ead{matt.visser@sms.vuw.ac.nz}

\address[sissa]{SISSA - International School for Advanced Studies, via Bonomea 265, 34136 Trieste, Italy.}
\address[infn]{INFN sezione di Trieste, via Valerio 2, Trieste, Italy.}
\address[vuw]{School of Mathematics and Statistics,
Victoria University of Wellington;
PO Box 600, Wellington 6140, New Zealand.}

\begin{abstract}
Analogue spacetimes can be used to probe and study physically interesting spacetime geometries by constructing, either theoretically or experimentally, some notion of an effective Lorentzian metric $[g_\mathrm{eff}(g,V,\,\Xi)]_{ab}$.  These effective metrics generically depend on some physical background metric $g_{ab}$, often flat Minkowski space $\eta_{ab}$, some ``medium'' with 4-velocity $V^a$, and possibly some additional  background fields $\Xi$. Electromagnetic analogue models date back to the 1920s, acoustic analogue models to the 1980s, and BEC-based analogues to the 1990s. The acoustic analogue models have perhaps the most rigorous mathematical formulation, and these acoustic analogue models really work best in the absence of vorticity, if the medium has an irrotational flow. 
This makes it difficult to model rotating astrophysical spacetimes, spacetimes with non-zero angular momentum, and in the current article we 
explore the extent to which one might hope to be able to model astrophysical spacetimes with angular momentum, (thereby implying vorticity in the 4-velocity of the medium). 

\smallskip\noindent
\emph{Date:} 14 February 2018; 19 February 2018, LaTeX-ed \today.

\smallskip\noindent
\emph{Preprinted as:}  arXiv: 1802.04785 [gr-qc]
%
\end{abstract}


\end{frontmatter}
\parindent0pt
\parskip7pt
\section{Introduction}\label{S:intro}

Analogue spacetimes have an almost century-long, complex, turbulent, and quite chequered history, see~\cite{Barcelo:2005,Visser:2007,Visser:2007b,Visser:2013}.
With hindsight Gordon's 1923 paper~\cite{Gordon:1923}, wherein he explored what would now be called electromagnetic analogues, was considerably more important and insightful than it may have seemed at the time.  The idea of electromagnetic analogue spacetimes --- precisely what distribution of permittivity $\epsilon$ and permeability $\mu$ (and possibly magneto-electric $\zeta$ effects) mimics a classical gravitational field --- subsequently became one of the  exercises in the Landau-Lifshitz volume on classical field theory~\cite{Landau-Lifshitz}. 
Scientific interest in these electromagnetic analogues is ongoing, see for instance~\cite{Plebanski:1960, Plebanski:1970, deFelice:1971, Joerg:2010, Joerg:2011a,Joerg:2011b, Schuster:2017, Schuster:2018, Schuster:2018b}, and references therein.

In 1981 Unruh developed the acoustic analogue spacetimes (subsequently called dumb holes)~\cite{Unruh:1981}, with further developments due to one of the present authors~\cite{Visser:1993,Visser:1997,Visser:2010}. While these early acoustic models were based on ordinary barotropic fluid mechanics, much subsequent work was based on the ``Madelung fluid'' interpretation of a quantum condensate wave-function --- typically a non-relativistic or relativistic BEC~\cite{Garay:1999,Garay:2000,Barcelo:2001,Barcelo:2003,Visser:2004,Visser:2005,Cropp:2016,Finke:2016,Braden:2017}.

\enlargethispage{30pt}
Such analogue models have been applied in mimicking several interesting spacetimes. However, angular momentum in the physical spacetime to be mimicked corresponds to vorticity in the flow of the medium used in setting up the analogue. See for instance~\cite{Visser:2004b,Richartz:2014,Cardoso:2016,Torres:2016,Torres:2017,Patrick:2018,Kerr-book,Kerr-intro,LTV-in-preparation}, and the more general background references~\cite{Barcelo:2005,Visser:2007,Visser:2007b,Visser:2013}.

 This raises the question of just how one might introduce vorticity into analogue spacetimes?
(This question is considerably trickier than one might naively think.) 
Certainly there is no difficulty at the level of ray optics or ray acoustics where the ``propagation cones" are quite flexible:
\begin{equation}
(g_\mathrm{eff})_{ab} \propto g_{ab} +[1-c_\mathrm{propagation}^{2}] V_a V_b.
\end{equation}
Here $V^a$ is the 4-velocity of the medium, while $c_\mathrm{propagation}$ is the propagation speed of whatever signal one is interested in. Because ``propagation cones" at best describe the effective metric up to an unknown and \emph{unknowable} conformal factor, in the ray optics or ray acoustics limit one can at best determine an effective metric up to proportionality~\cite{Barcelo:2005,Visser:2007,Visser:2007b,Visser:2013}.
(In this eikonal limit see also~\cite{Leonhardt:1999}.)
However, significant subtleties arise when one wishes to derive a wave equation suitable for wave optics or wave acoustics~\cite{Barcelo:2005,Visser:2007,Visser:2007b,Visser:2013}. 

For instance, the simplest of the non-relativistic~\cite{Unruh:1981,Visser:1993,Visser:1997}, and relativistic~\cite{Visser:2010} acoustic models, (which is where we have the best mathematical control over the formalism), were explicitly constructed to be vorticity-free. 
In contrast the non-relativistic vorticity-supporting acoustic model developed in~\cite{Perez-Bergliaffa:2001} was somewhat more complex, requiring the use of Clebsch potentials. 
A more recent 2015 model~\cite{Cropp:2015,Giacomelli:2017} used charged BECs (both non-relativistic and relativistic, in the usual acoustic limit of the Gross--Pitaveskii equation, where the quantum potential is neglected).

That one might strongly desire to add vorticity to a wide class of analogue spacetimes is driven by both theoretical and experimental issues: Certainly sound propagates on physical vortex flows, and it would be highly desirable to have a suitable well-controlled mathematically precise wave-equation that goes beyond the Pierce approximation~\cite{Perez-Bergliaffa:2001,Pierce}.\footnote{As we shall see later on, the Pierce approximation allows one to derive
an approximate wave equation for sound propagation in an inhomogeneous fluid by
assuming that the characteristic length and time scales for the ambient
medium are larger than the corresponding scales for the acoustic disturbance.
Under these conditions the system can be described by a wave equation that is
correct to first order in the derivatives of the ambient quantities.} 

Certainly the Kerr congruence in a rotating Kerr black hole has non-zero vorticity~\cite{Visser:2004b,Richartz:2014,Cardoso:2016,Torres:2016,Torres:2017,Patrick:2018,Kerr-book,Kerr-intro,LTV-in-preparation}, so any analogue model of the Kerr spacetime (or any spacetime with nonzero angular momentum) will need to include some notion of vorticity.
In view of these comments, a  key point of this letter would be to better understand the notion of vorticity in the charged BEC models~\cite{Cropp:2015,Giacomelli:2017} and in Gordon's 1923 model~\cite{Gordon:1923}.

\section{The Gordon metric}  

Gordon's 1923 effective metric is:
\begin{equation}
(g_\mathrm{eff})_{ab} = \eta_{ab} +[1-n^{-2}] V_a V_b.
\end{equation}
In Gordon's original article, (a condensed-matter model), 
the refractive index $n$ (and $\epsilon$ and $\mu$) and the 4-velocity of the medium $V^a$ were {mainly} (but not always) taken to be position-independent constants. The refractive index $n$ (and $\epsilon$ and $\mu$) were always taken to be isotropic.
Furthermore Gordon was {often} working in the ray optics (eikonal) limit, and certainly assuming a background flat Minkowski space. Somewhat oddly, Gordon did not seem to recognize the need for the now well-known consistency condition: $\epsilon=\mu=n$, (plus additionally setting the magneto-electric effect $\zeta$ to zero)~\cite{Schuster:2017, Schuster:2018,Schuster:2018b}. 

Various generalizations to Gordon's model that one might consider would include:
(1) Fully general position-dependent $n(x)$ and $V^a(x)$; with considerable hindsight this is implicit but not really explicit in Gordon's 1923 article. 
(2) Introducing non-zero vorticity for the background flow.
(3) Going beyond the eikonal (ray) approximation to consider wave physics.
(4) Introducing a non-trivial background metric.

\section{{The Unruh metric}}

 The Unruh 1981 metric for acoustic perturbations in an irrotational barotropic inviscid fluid can be given in ADM-like form~\cite{Unruh:1981}:
\begin{equation}
(g_\mathrm{eff})_{ab} = {\rho_0\over c_s}\left[\begin{array}{c|c}-[c_s^{2}-v^2]&-v_j\\  \hline 
-v_i& \delta_{ij}  \end{array}\right]; 
\end{equation}
Here $v_i$ is the 3-velocity of the fluid, $\rho_0$ is its density, and $c_s$ is the speed of sound. The irrotational condition $\nabla\times v = 0$ (and hence $v = \nabla \Phi$) is built in as the 
very first equation of~\cite{Unruh:1981}. (This vorticity-free assumption continues to hold in~\cite{Visser:1993,Visser:1997}.)

\section{{Extensions beyond Unruh 1981}}
Various partial extensions of the acoustic analogue models include:
\subsection{{Relativistic barotropic irrotational acoustics}}
The metric here is best presented in Gordon-like form~\cite{Visser:2010}
\begin{equation}
(g_\mathrm{eff})_{ab} = \Omega^2\left( g_{ab} +[1-c_s^{2}] V_a V_b \right); 
\end{equation}
\begin{equation}
V_a = {\nabla_a \Phi\over||\nabla\Phi||}.
\end{equation}
Here $c_s$ is the speed of sound, while $g_{ab}$ is an arbitrary background metric. 
The conformal factor $\Omega$ is a specific known but somewhat messy function of the barotropic equation of state $\rho =  \rho(p)$. The only significant physics restriction on this metric is that the background 4-velocity $V_a$ is irrotational, that is {vorticity-free}, in the relativistic sense $V_{[a} V_{b,c]} = 0$.

\subsection{{Non-relativistic barotropic rotational acoustics}}
  
 The central idea here is to introduce vorticity through the use of Clebsch potentials.
 Any 3-vector field can be put in the form
 \begin{equation}
v = \nabla \phi + \beta \nabla \gamma.
\end{equation}
(See~\cite{Perez-Bergliaffa:2001} and references therein.) One then decomposes the fluid 3-velocity into background plus (small) perturbation
\begin{equation}
v = v_0 + v_1;   \qquad v_1 = \nabla \psi + \xi;
\end{equation}
with
\begin{equation}
\psi =   \phi_1 + \beta_0\gamma_1; \qquad 
\xi = \beta_1\nabla\gamma_0-\gamma_1 \nabla\beta_0.
\end{equation}
The analysis then leads to a \emph{system} of PDEs. 
The effective metric is again ADM-like form
\begin{equation}
(g_\mathrm{eff})_{ab} = {\rho_0\over c_s}\left[\begin{array}{c|c}-[c_s^{2}-v^2]&-v_j\\  \hline 
-v_i& \delta_{ij}  \end{array}\right]; 
\end{equation}
with inverse
\begin{equation}
(g_\mathrm{eff}^{-1})^{ab} = {1\over \rho_0c_s}\left[\begin{array}{c|c}-1&-v^j\\  \hline
-v^i& c_s^2\delta_{ij}  - v^i v^j\end{array}\right].
\end{equation}
(Spatial indices are raised and lowered using the flat Euclidean metric $\delta_{ij}$.)
Then
\begin{equation}
\Box_g \psi =  -{c_s\over\rho_0^2}  \nabla\cdot(\rho_0 \xi),
\end{equation}
where $\Box_g$ is the d'Alembertian wave operator for the metric $(g_\mathrm{eff})_{ab}$, whereas in terms of the advective derivative
\begin{equation}
{d\xi\over dt} = \nabla\psi\times \omega_0 - (\xi\cdot\nabla) v_0.
\end{equation}
Thus the rotational part of the velocity perturbation, $\xi$, acts as a source for the wave equation for $\psi$. Conversely the background vorticity, $\omega_0$ helps drive the 
evolution of the rotational part of the velocity perturbation, $\xi$.

In this context, Pierce's approximation~\cite{Pierce} amounts to asserting that the background velocity gradients $\nabla v_0$ be much smaller than the frequency of the wave, which automatically implies that background  vorticity $\omega_0$ is much smaller than the frequency of the wave.
Under those conditions Pierce argues that {both} the rotational part of the velocity perturbation  {and its gradient,
(both $\xi$ and $\nabla \xi$),} will always remain small and can safely be neglected~\cite{Perez-Bergliaffa:2001, Pierce}. Under those circumstances one obtains an \emph{approximate} wave equation
\begin{equation}
\Box_g \psi \approx 0. 
\end{equation}
Generally though, one simply has to keep track of the extra complications coming from $\xi$, the rotational part of the velocity perturbation. 

In summary, this analogue model certainly exhibits an effective metric, and the effective metric certainly can support background vorticity, but the model is ``contaminated'' by the presence of the extra field $\xi$, which complicates any attempt at setting up a clean ``fully geometric'' interpretation.

\subsection{{Charged non-relativistic BECs}}

The metric here is given in ADM-like form~\cite{Cropp:2015,Giacomelli:2017}:
\begin{equation}
(g_\mathrm{eff})_{ab} = {\rho_0\over c_s}\left[\begin{array}{c|c}-[c_s^{2}-v^2]&v_j\\  \hline v_i& \delta_{ij}  \end{array}\right]; 
\end{equation}
with the 3-velocity of the effective Madelung fluid defined by
\begin{equation}
v_i = \nabla_i \Phi-{eA_i\over mc}.
\end{equation}
The BEC wavefunction is $\Psi = \sqrt{\rho_0}\;  e^{i\Phi} = ||\Psi|| e^{i\Phi}$, while $c_s$ is the speed of sound in the condensate, and the purely spatial 3-vector $v \propto \nabla\Phi-eA$ is gauge invariant. The key point is that $\nabla_i \Phi-eA_i$ makes sense only if one has a condensate.
This would in principle seem to allow the background flow to have some vorticity while keeping the perturbations irrotational.   Specifically the vorticity would be
\begin{equation}
\omega = {e B\over m c}.
\end{equation}

Unfortunately one also has $v \propto j_\mathrm{London}$, the so-called London current that is central to the analysis of the Meissner effect. 
Indeed, there is widespread agreement within the condensed matter community that \emph{any} charged BEC, (not just a BCS superconductor, where formation of the Cooper pairs, and condensation of the Cooper pairs are essentially simultaneous), will exhibit the Meissner effect --- magnetic flux expulsion.
See for instance~\cite{Meissner:2003,Meissner:1951,Meissner:1955}. 
This would naively seem to confine any vorticity to a thin layer of thickness comparable to the London penetration depth.
However there is a trade-off between the healing length (which controls the extent to which one can trust the effective metric picture) and the London penetration depth. 

Let us be more quantitative about this:
The London penetration depth and healing length are  given by:  
\begin{equation}
\lambda = \sqrt{m \over \mu_0 n q^2}; \qquad \xi = {1\over\sqrt{8\pi n a}}.
\end{equation}
Here $m$ is the mass of the atoms making up the charged BEC, $\mu_0$ is the magnetic permeability in vacuum, $n$ is the number density of atoms in the condensate; $q=Qe$ is the charge of each atom, and $a$ is the scattering length.
In particular, for the ratio of penetration depth to healing length the number density $n$ cancels and  using $\epsilon_0\mu_0= 1/c^2$ we have
\begin{eqnarray}
{\lambda\over \xi} &=& \sqrt{8\pi a m\over\mu_0 q^2}= \sqrt{8\pi \epsilon_0 c^2 a m\over q^2}.
\end{eqnarray}
Write $q=Qe$ and $m=Nm_p$, where $N$ is the atomic mass number.
Then
\begin{equation}
{\lambda\over \xi} =  {\sqrt{N}\over Q} \sqrt{2 am_p c\over \alpha \hbar}.
\end{equation}
Then in terms of the Bohr radius, $a_0= \alpha^{-1} \hbar/(m_e c)$, one has
\begin{equation}
{\lambda\over \xi} =  {\sqrt{N}\over Q} {1\over\alpha} {\sqrt{2 m_p\over m_e}}\sqrt{a\over a_0}.
\end{equation}

We can in principle make this ratio large, simply by tuning to a Feshbach resonance.
Let us write $a= (a/a_*)\, a_*$, where $a_*$ is the zero-field scattering length before tuning to a Feshbach resonance.
We know that $\sqrt{2 m_p/ m_e}\approx 60$.  For a heavy atom charged BEC $\sqrt{N}/Q \approx 9$ and typically $a_* \approx 100 a_0$.  Then
\begin{equation}
{\lambda\over \xi} \approx  750000 \sqrt{a\over a_*}.
\end{equation}
So there is a significant separation of scales between healing length and London penetration depth, which can be made even larger by tuning to a Feshbach resonance.

The net outcome of this discussion is that despite potential problems due to the Meissner effect there is a parameter regime in which we can simultaneously have vorticity penetrate deep into the bulk and still trust the effective metric formalism.

Another interesting feature of this non-relativistic construction (not commented on previously) is that even if the background has vorticity, the perturbations are vorticity-free. This is a side-effect of the Madelung representation, (and the approximation of neglecting the quantum potential). One again takes 
\begin{eqnarray}
\psi_\mathrm{total} &=& \Psi \, (1+\psi_\mathrm{perturbation})
\nonumber\\
&=&  \sqrt{\rho_0} \, e^{i\Phi} \,(1+\psi_\mathrm{perturbation}),
\end{eqnarray}
and obtains a d'Alembertian PDE for $\psi_\mathrm{perturbation}$.
So in contrast to the previous model, no extra fields need to be introduced.

\subsection{{Charged relativistic BECs}}
The metric here is best presented in Gordon-like form~\cite{Cropp:2015,Giacomelli:2017}:
\begin{equation}
(g_\mathrm{eff})_{ab} = {\rho_0\over c_s}\left( \eta_{ab} +[1-c_s^{2}] V_a V_b \right); 
\end{equation}
with
\begin{equation}
V_a = {\nabla_a \Phi-eA_a\over||\nabla\Phi-eA||}.
\end{equation}
It is now the RBEC wavefunction that is written in terms of the Madelung representation $\Psi = \sqrt{\rho_0}\;  e^{i\Phi}$, while $c_s$ is the speed of sound in the condensate, and the 4-velocity $V \propto \nabla\Phi-eA$ is gauge invariant. This (formally) allows the background flow to have some vorticity while keeping the perturbations irrotational.   Specifically for the 4-vorticity we have
\begin{equation}
\epsilon_{abcd} \omega^d = 
V_{[a} V_{b,c]}  = e\; {V_{[a} F_{bc]} \over ||\nabla\Phi-eA||}.
\end{equation}
Working in the rest frame of the fluid we see
\begin{equation}
||\omega|| = {e\; ||B|| \over ||\nabla\Phi-eA||}.
\end{equation}
The key point is that $\nabla_a \Phi-eA_a$ is a gauge invariant 4-vector field that makes sense only if one has a condensate.

The same  potentially problematic issue regarding the Meissner effect also arises in this relativistic setting. The 4-velocity now satisfies $V\propto \nabla\Phi-eA \propto J_\mathrm{London}$, where this is now the London 4-current $J_\mathrm{London} = (\rho_\mathrm{London}, j_\mathrm{London})$. Naively, the magnetic field (and hence the vorticity) will be confined to a thin transition layer  of thickness comparable to the London penetration depth. 
However the same parameter regime as was considered for the non-relativistic case will still apply in the full relativistic setting: Drive the London penetration depth large while holding the healing length constant.

\section{{Extensions beyond Gordon 1923}}
\newcommand{\pdet}{\ensuremath{\mathrm{pdet}}} 
For linear constitutive $\epsilon$-$\mu$-$\zeta$ electrodynamics
a recent fundamental result for the effective metric is
\begin{eqnarray}
(g_\mathrm{eff})_{ab} &=& 
-\sqrt{\frac{-\det(g^{\bullet\bullet})}{\pdet(\epsilon^{\bullet\bullet})}}\, V_a V_b 
\nonumber\\
&&
+ \sqrt{\frac{\pdet(\epsilon^{\bullet\bullet})}{-\det(g^{\bullet\bullet})}} [\epsilon^{\bullet\bullet}]^\#_{ab},
\end{eqnarray}
subject to the standard consistency constraint
\begin{equation}
\epsilon^{ab}=\mu^{ab};  \qquad \zeta^{ab}=0.
\end{equation}
(Note specifically the pseudo-determinant~\cite{Schuster:2017} and Moore--Penrose pseudo-inverse appearing above.\footnote{The pseudo-determinant $\pdet(X)$ is simply the product over non-zero eigenvalues. For symmetric matrices $X$, the Moore--Penrose pseudo-inverse $X^\#$ simplifies to diagonalizing the matrix, taking the reciprocal of the non-zero eigenvalues, and then undoing the diagonalization. See for instance~\cite{Schuster:2017} and references therein.})
Here $V^a$ is the 4-velocity of the medium, and one has generalized Gordon 1923 by allowing for non-trivial permittivity and permeability tensors. These are both symmetric and transverse in the sense that $\epsilon^{ab}V_b = 0 = \mu^{ab} V_b$, so that in the rest frame of the medium they reduce to $3\times3$ symmetric tensors.
The formalism was carefully set up so that there is no constraint on the background 4-velocity;
 it can in principle be arbitrary~\cite{Schuster:2017}. 
The formalism was also carefully set up so that there is no constraint on the background 4-geometry;
 it can in principle be arbitrary~\cite{Schuster:2017}. 
 
\medskip
\noindent
If we now assume an isotropic medium, then
\begin{equation}
\epsilon_{ab} = \epsilon (g_{ab}+V_a V_b);
\end{equation}
\begin{equation}
\mu_{ab} = \mu (g_{ab}+V_a V_b);
\end{equation}
\begin{equation}
\zeta_{ab} = \zeta (g_{ab}+V_a V_b).
\end{equation}
The compatibility condition then reduces to
\begin{equation}
\epsilon = \mu = n; \qquad \zeta =0;
\end{equation}
so that
\begin{equation}
(g_\mathrm{eff})_{ab} =  n^{3/2} V_a V_b + n^{1/2}(g_{ab}+V_aV_b).
\end{equation}
That is:
\begin{equation}
(g_\mathrm{eff})_{ab} =  \sqrt{n} \left\{ g_{ab} + [1-n^{-2}] V_a V_b\right\}
\end{equation}
The $\sqrt{n}$ pre-factor is completely conventional; its presence is simply due to the conformal invariance of electromagnetism in (3+1) dimensions and the convenient demand that 
$\det[(g_\mathrm{eff})_{ab}] = \det[g_{ab}]$, see~\cite{Schuster:2017}. 

\noindent
We could just as well write
\begin{equation}
(g_\mathrm{eff})_{ab} =  \Omega^2 \left\{ g_{ab} + [1-n^{-2}] V_a V_b\right\}.
\end{equation}
The conformal factor $\Omega^2$  is arbitrary, the background metric $g_{ab}$ is arbitrary, the refractive index $n$ is also arbitrary (subject only to the consistency condition $\epsilon = \mu = n$), and finally the 4-velocity $V_a$ is arbitrary. Certainly in principle any arbitrary non-zero background vorticity is allowed. This pretty much settles the main physics issue --- there is no deep physics obstruction to putting vorticity into the Gordon metric at the level of wave optics.

\section{Discussion}

While, as we have seen, introducing vorticity into analogue models at the level of ray optics or ray acoustics (the eikonal limit) is straightforward, the situation at the level of wave optics or wave acoustics is considerably more subtle. Fortunately, for wave optics the Gordon metric (now suitably generalized and placed in a more up-to-date context) provides a suitable model.

There are of course many additional relevant articles on related topics from within the astrophysics, condensed matter, and optics communities. See for instance~\cite{Leonhardt:1999} and the extensive list of references in~\cite{Barcelo:2005}.  We have unavoidably had to be somewhat selective in our selection of references. Relatively recent developments include the notions  of   ``quantum vorticity''~\cite{Good:2014,Xiong:2014,Huang:2015} and  ``holographic vorticity''~\cite{Caldarelli:2011,Leigh:2011,Leigh:2012,Eling:2013}.

Taken as a whole, these observations collectively give us confidence that it is likely to be possible to mimic the Kerr solution at the wave optics or wave acoustics level --- presumably through some ``Kerr-Gordon'' form of the metric. It is already known that the Schwarzschild metric can be put into Gordon form~\cite{Giacomelli:2017,Rosquist}:
\begin{equation}
g_{ab} =\sqrt{n} \left( \eta_{ab} + [1-n^{-2}]V_a V_b\right); 
\end{equation}
\begin{equation}  
V_a = \left(- \sqrt{1+2m/r}; \sqrt{2m/r} \; \hat r_i\right).
\end{equation}
Here $n$ is an arbitrary constant, $V_a$ is a 4-velocity, and the parameter $m$ is proportional to the physical mass.
The overall conformal factor $\sqrt{n}$ in the metric enforces $\det(g)=-1$. 
It is easy to check that this metric this is Ricci flat. A similar ``Kerr-Gordon'' construction for the Kerr spacetime would be very interesting~\cite{LTV-in-preparation}, and looking for such a construction is largely the reason we became interested in the ideas presented in the current article.

\vspace{-10pt}

\section*{Acknowledgements}
The authors wish to thank Andrea Trombettoni for useful comments on the Meissner effect in charged BECs.\\
SS was supported by a VUW (Victoria University of Wellington) PhD Scholarship.\\
MV was supported by the Marsden Fund, which is administered by the Royal Society of New Zealand.


\end{document}